\documentclass{aa}  
\usepackage{graphicx}
\usepackage{txfonts}
\usepackage{color}

\def \pmbmath{\mathpalette\pmbmathaux}
\def \pmbmathaux#1#2{
         \pmbtext{$#1#2$}}
\def \pmbtext#1{\leavevmode
     \setbox0\hbox{#1}
     \kern-0,2pt \copy0 \kern-\wd0
     \kern0,4pt \copy0 \kern-\wd0
     \kern-0,2pt \raise0,3pt \box0}

\begin{document}
\title{An anisotropic turbulent model for solar coronal heating}
\author{B. Bigot \inst{1,2}, S. Galtier\inst{1} \and H. Politano\inst{2}}
\institute{Institut d'Astrophysique Spatiale, B\^atiment 121, Universit\'e Paris-Sud XI, UMR 8617, 
91405 Orsay, France \\
\and
Universit\'e de Nice-Sophia Antipolis, CNRS UMR 6202, Observatoire de la C\^ote d'Azur, 
BP 42229, 06304 Nice Cedex 4, France\\
\email{bbigot@obs-nice.fr}}
\abstract
{We present a self-consistent model of solar coronal heating
in which we include the dynamical effect of the background magnetic field along 
a coronal structure by using exact results from wave MHD turbulence.}
{We evaluate the heating rate and the microturbulent velocity for comparison with observations in 
the quiet corona, active regions and also coronal holes.}
{The coronal structures are assumed to be in a turbulent state maintained by the slow erratic motion  
of the magnetic footpoints. A description of the large-scale and the unresolved small-scale dynamics 
are given separately. From the latter, we compute exactly (or numerically for coronal holes) turbulent 
viscosites used in the former to self-consistently close the system and derive the 
heating flux expression.}
{We show that the heating rate and the turbulent velocity compare favorably with coronal observations.}
{Although the Alfv\'en wave turbulence regime is strongly anisotropic, and could reduce a priori the 
heating  efficiency, it provides a unexpected satisfactory model of coronal heating for both magnetic 
loops and open magnetic field lines.} 
\maketitle

\section{Introduction}

Information about the solar corona from spacecraft missions like Yohkoh, SoHO (Solar 
\& Heliospheric Observatory) or TRACE (Transition Region And Coronal Explorer) launched in 
the 1990s reveals a very dynamical and complex medium 
structured in a network of magnetic field lines. These observations clearly 
demonstrated the fundamental role of the magnetic field on the plasma dynamics in the solar 
atmosphere. The solar corona contains a variety of structures over a broad range of scales 
from about $10^5$ km until the limit of resolution (about one arcsec). It is very likely that 
structures at much smaller scales exist but have not yet been detected. The new spacecrafts 
STEREO, Hinode or SDO (Solar Dynamics Observatory) will help our understanding 
of the small-scale nature of the corona. 

Observations in UV and X-ray show a solar corona that is extremely hot with temperatures 
exceeding $10^6$ K -- close to hundred times the solar surface temperature. These 
coronal temperatures are highly inhomogeneous: in the quiet corona much of the plasma lies 
near $1$--$2 \times 10^6$ K and $1$--$8 \times 10^6$ K in active regions. Then, one of the 
major questions in solar physics concerns the origin of such high values of coronal temperature. 
The energy available in the photosphere is clearly sufficient to supply the total coronal losses 
(Withbroe \& Noyes 1977) which is estimated to be $10^4 {\rm J \, m}^{-2} {\rm s}^{-1}$ for active 
regions and about one or two orders of magnitude smaller for the quiet corona and coronal holes. 
The main issue is thus to understand how the available photospheric energy is transferred and 
accumulated in the solar corona, and by what processes it is dissipated. 

It is widely believed that the energy input comes from the slow random motion of the convective 
layer below the photosphere, but the mechanisms that heat the solar corona remain controversial. 
Heating models are often classified into two categories: ``AC'' and ``DC'' heating. 
``AC'' heating by Alfv\'en waves was suggested for the first time by Alfv\'en (1947). 
In non-uniform plasmas, this heating mechanism could be sustained by the resonant absorption 
of Alfv\'en waves (Hollweg 1984), or by phase mixing (Heyvaerts \& Priest 1983). 
The necessary condition for ``AC'' heating is that the characteristic time of motion excitation
has to be shorter than the characteristic time of wave propagation across the coronal loops. 
In the opposite situation, the most efficient heating is the ``DC'' heating by direct current. In this case, 
the quasi-static energy input allows the accumulation of magnetic energy, the generation 
of currents and finally heating, for example, by reconnection of magnetic field lines 
(\cite{Priest00}) or by resistive dissipation of current sheets. However, in this simple version, these mechanisms are not efficient enough to explain coronal heating. Additional processes 
are thus generally introduced like turbulence, which can generate 
small scales where dissipation is much more efficient (see {\it e.g.} \cite{Gomez88},\cite{Gomez92}, 
\cite{HP92,Einaudi96},\cite{Dmitruck97}, \cite{Galtier98}, \cite{Galtier03}, and \cite{Buchlin07}). 
Moreover, turbulence could explain the measurements of nonthermal velocities revealed by the width 
of EUV and FUV lines. Observations of the transition region and corona 
of the quiet Sun from SUMER onboard SoHO reveal nonthermal velocities of about 
$30 {\rm km \, s}^{-1}$ for temperatures 
around $3 \times 10^{5} \rm{K}$ with a peak up to $55 \rm{km \, s}^{-1}$ for some S IV lines 
(\cite{Warren97}; \cite{Chae98}).

\begin{figure}[ht]
\centering
\includegraphics[width=9.cm]{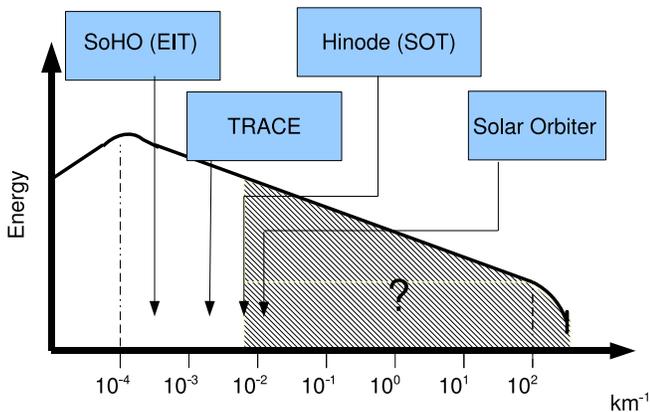}
\caption{Schematic view of the energy spectrum of the coronal plasma in logarithmic coordinates. 
The spatial resolution of different spacecraft instruments is reported: EIT/SoHO ($1850$ km), 
TRACE ($350$ km), SOT/Hinode ($150$ km) and Solar Orbiter ($35$ km). The gray zone indicates 
currently unresolved small-scales which will be modeled by turbulent viscosities.}
\label{Intro_Heat}
\end{figure}

Many observations of the solar atmosphere tend to show plasma in a turbulent state 
with a Reynolds number evaluated at about $10^{12}$. In particular, the most 
recent Hinode pictures seem to show a magnetic field controlled by plasma turbulence 
at all scales (Nature 2007; see, also, \cite{Doschek07}). Thus, the turbulent activity of the 
corona is one of the key issues to understand the heating processes. 
In the framework of turbulence, the energy supplied by the photospheric motion 
and transported by Alfv\'en waves through the corona is transferred towards 
smaller and smaller scales by nonlinear coupling between modes (the so-called 
energy cascade) until dissipative scales are reached from
which the energy is converted into heating. 
The main coronal structures considered in such a scenario are the magnetic loops which 
cover the solar surface, in active and quiet regions. 
Each loop is basically an anisotropic bipolar structure anchored in the photosphere. It 
forms a tube -- or an arcade -- of magnetic fields in which the dense and hot matter is confined. 
Because a strong guiding magnetic field ($\mathbf{B_0}$) is present, the nonlinear cascade 
that occurs is 
strongly anisotropic with small scales mainly developed in the $\mathbf{B_0}$ transverse planes. 
In Figure \ref{Intro_Heat}, we present a schematic view of the turbulent energy spectrum expected 
for the coronal plasma which follows a power law over several decades. The larger scales may be 
determined directly by observation and correspond to about $10^5$ km. The inertial range -- that 
determines the range of scales where the turbulent cascade operates -- roughly starts between 
$10^4$ km  and $10^3$ km, and extends down to unresolved (by spacecraft observations) scales 
which may be estimated, from dimensional analysis, as of a few 
meters. The spatial resolution of different instruments is also reported to show the gap 
that we need to fill in order to completely resolve the heating processes that occur at the smallest
dissipative scales.

\begin{figure}[ht]
\centering
\includegraphics[width=9.cm]{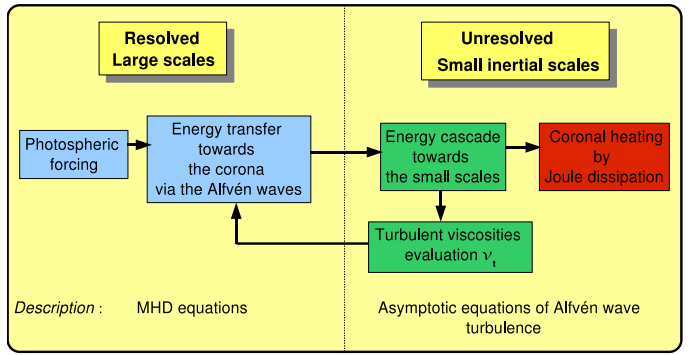}
\caption{Schematic view of the self-consistent heating model. The resolved large-scales coronal flow is 
described by the "large-scale" MHD equations where turbulent (instead of standard) viscosity and 
resistivity are used 
to determine the inertial small-scale nonlinear dynamics. From these equations we compute the energy 
flux function released by the photospheric motion. The energy is transported by Alfv\'en waves through 
the corona and the nonlinear interactions between wave packets generate an inertial cascade, {\it ie.} a 
production of smaller and smaller scales, that finally reach the dissipative smallest scales from which the 
energy is converted into ohmic dissipation. The unresolved inertial small-scales are described by the asymptotic equations of Alfv\'en wave turbulence from which the turbulent transport coefficients are 
computed. A self-consistent model is finally obtained by introducing the expression of these turbulent 
viscosities into the energy flux function. Then, the energy flux function is entirely determined and the 
heating rate can be evaluated.}
\label{ShemaHeating}
\end{figure}

The aim of the present study is coronal heating to perform of modeling by magnetohydrodynamic (MHD) turbulence. 
We explain the connection between the large scales
(at which energy is injected at loop footpoints through photospheric motion) and the smallest scales 
(at which  energy is dissipated and converted into heat).
The input energy propagates through the 
network of magnetic field lines by "inward" and "outward" Alfv\'en waves which nonlinearly interact 
and produce a turbulent cascade towards small scales. The foundation of our model is the one 
originally proposed by Heyvaerts \& Priest (1992) where the unresolved small-scales are modeled 
through turbulent viscosities. These viscosities were extracted from an {\it ad hoc} EDQNM (Eddy-Damped 
Quasi-Normal Markovian) closure model of MHD turbulence developed in spectral space 
for isotropic flows (\cite{Pouq76}). Spacecraft missions like SoHO, TRACE, or more recently Hinode show that this 
assumption is clearly not adapted to magnetic loops which are characterized by a strong longitudinal 
mean field ${\mathbf B_0}$, compared to the magnetic perpendicular components, whose anisotropic nonlinear 
effects on the turbulent plasmas dynamics are more than likely, as many numerical simulations have shown. 
For that reason, in the present model, we use turbulent viscosities computed from an asymptotic (exact) 
closure model of MHD turbulence (\cite{Galtier00,Galtier02}), also called Alfv\'en wave 
turbulence. 
Signatures of such a regime have been detected in the middle magnetosphere of Jupiter (Saur et al., 2002). 
In our turbulent heating model, the unresolved small-scale equations are perturbatively developed  
around a strong magnetic field $\mathbf{B_0}$ which leads to a strongly anisotropic turbulence.
From the derived equations, we compute the turbulent viscosities which eventually allow us to 
obtain a self-consistent free-parameter model of coronal heating from which we predict a heating rate and a 
turbulent velocity that favorably compare with observations. 
Figure \ref{ShemaHeating} summarizes the schematic algorithm followed to obtain a 
self-consistent heating model.

The organization of the paper is as follows: the next Section is dedicated to the large-scale description 
of magnetic loops where, in particular, we rederive some main results 
obtained by Heyvaerts \& Priest (1992), using a more adapted notation. 
Then, in Section \ref{sectionSS}, the small-scale description is given and the turbulent viscosities are obtained. 
The model predictions (heating rate and turbulent velocity) for magnetic loops in active 
regions and in the quiet corona are given in Section \ref{chauffage1}. They are generalized to open 
magnetic lines in coronal holes in Section \ref{chauffage2}. 
The last Section is devoted to the discussion and conclusion.

\section{Large-scale description of magnetic loops}
\subsection{Geometry and boundary conditions}

\begin{figure}
\includegraphics[width=8.5cm]{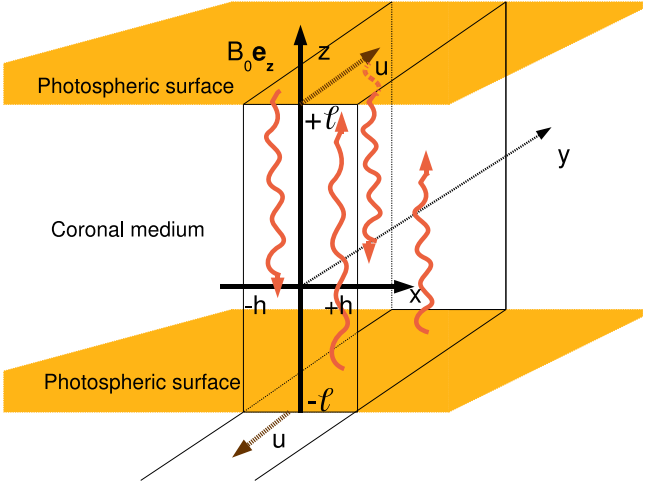}
\caption{Arcade geometry: the coronal plasma is confined in a volume delimited by $-h<x<+h$, 
$-\infty<y<+\infty$ and $-\ell< z <+\ell$. The photosphere appears as two boundary planes at 
altitudes $z=-\ell$ and $z=+\ell$. Note the presence of Alfv\'en waves propagating in opposite 
directions along $\mathbf{B_0}$.}
\label{LoopModel}
\end{figure}

We consider a set of magnetic loops -- an arcade -- anchored in the photosphere and 
subjected to the erratic motion of convective cells. For simplicity, this arcade 
has a rectangular cross surface which defines the perpendicular directions and it is elongated 
along the longitudinal magnetic field $\mathbf{B_0}=B_0 \mathbf{e_z}$ that defines the parallel 
direction as shown in Figure \ref{LoopModel}. This magnetic structure is filled by a plasma which 
will be modeled as an incompressible MHD fluid. In the first part of this paper, we are concerned 
with magnetic loops; later on we will extend our results to open magnetic field lines for which the 
boundary conditions will be modified. For a magnetic arcade, the boundary limits are made of two 
planes at altitudes $z=-\ell$ and $z=+\ell$ which are connected by the uniform magnetic field 
$\mathbf{B_0}$. 
Thus, the length of the loop in the longitudinal direction is $2 \ell$. The cross surface is delimited as 
$-h<x<+h$ and $-\infty<y<+\infty$. In other words, we assume an arcade characterized by a 
thickness of $2h$, where $h$ is the coherence length of the magnetic field at
the base of the corona, and with a large depth compare to the other length scales, with a translational 
invariance assumed in the $y$-direction. Note that we do 
not treat the expansion of the magnetic field at the footpoint loops where $B_0$ has to be seen 
as an average value of the field at the base of the corona rather than in the photosphere itself. 

The energy reservoir for the solar corona resides in the sub-photospheric convective layer. 
Indeed, the high-$\beta$ plasma 
in the photosphere constrains the magnetic field lines to be convected by the plasma flow. 
These motions eventually induce a shearing and a torsion of coronal magnetic field lines. 
The boundary motions at footpoint loops are assumed to be perpendicular and independent of 
the coronal plasma. More precisely, we assume for the boundary velocity field parallel to the y-axis:
\begin{equation}
\mathbf{u}(x,z=\ell) =  -\mathbf{u}(x,z=-\ell) = U(x)\mathbf{e_y}   
\label{BoundCI_vel} \ .
\end{equation} 
This input of energy is then 
propagated through the coronal medium where the low-$\beta$ confines the plasma into the 
magnetic arcade. These boundary motions are then able to produce and maintain a state of 
full turbulence in the corona. Note finally the absence of mass flow through the boundaries.

\subsection{The MHD model}
\label{mhdmodel}

The standard incompressible MHD equations are used as a first approximation to 
model the solar plasma, namely:
\begin{eqnarray}
\partial_t \mathbf{u} + \mathbf{u}\cdot\mathbf{\nabla} \mathbf{u}   &=&
- \mathbf{\nabla}P/\rho_0 + \mathbf{j}\times\mathbf{B}/\rho_0 + \nu \Delta  \mathbf{u} ,
\label{mhd1}\\
\mathbf{\nabla} \cdot \mathbf{u}  &=&  0 , 
\label{divu}\\
\partial_t \mathbf{B}  &=& \mathbf{\nabla}\times(\mathbf{u} \times\mathbf{B}) 
+ \eta \Delta \mathbf{B} , 
\label{mhd2}\\
\mathbf{\nabla} \cdot \mathbf{B} &=& 0 , 
\label{divb}
\end{eqnarray}
\begin{equation}
\partial_t \left(\rho_0 \frac{u^2}{2} + \frac{B^2}{2\mu_0} \right) +
\mathbf{\nabla} \cdot \mathbf{\mathcal {F}} = \mathcal{S} - \mathcal{D},
\label{EnergyEq}
\end{equation}
\begin{equation}
\mathbf{\mathcal{F}}   = \mathbf{u}\cdot\left(\rho_0 \frac{u^2}{2} + P - \bar{\bar{\sigma}}\right)
+ \frac{\mathbf{E}\times\mathbf{B}}{\mu_0},
\label{Feq}
\end{equation}
\begin{equation}
\mathbf{E}  + \mathbf{u} \times \mathbf{B} =\mu_0 \eta \mathbf{j} , 
\label{ohm}
\end{equation}
where $\mathbf{u}$ is the velocity, $\mathbf{B}$ the magnetic field (such as $\mathbf{B}=\mathbf{B_0} 
+\mathbf{b}$), $\mathbf{j}=\mathbf{\nabla}\times \mathbf{b}/\mu_0$ the current density
(where $\mu_0$ is the permeability of vacuum), $\mathbf{E}$ the electric field, $\bar{\bar{\sigma}}$ the 
viscous stress tensor (with $\sigma_{ij}=\rho_0 \nu (\partial u_i/\partial r_j + \partial u_j/\partial r_i))$,  
$\nu$ the kinematic viscosity, $\eta$ the magnetic diffusivity and $\rho_0$ the uniform mass density, 
and $P$ is the gas pressure. In the energy equation (\ref{EnergyEq}), the right hand side term $\mathcal{S}$ 
is the large-scale energy input created by the photospheric motion and $\mathcal{D}$ is the 
dissipation which mainly happens at the smallest scales because of, {\it e.g.}, viscous and resistive effects. 
The left hand side term, $\mathbf{\nabla} \cdot \mathbf{\mathcal {F}}$, denotes the nonlinear transfer of 
total (kinetic plus magnetic) energy. (Note that all terms in equation (\ref{EnergyEq}) are expressed in 
unit volume.) Equation (\ref{Feq}) describes how the total energy is nonlinearly transferred at different 
scales by velocity transport, by work due to the gas pressure, the viscous stress  or the 
electromagnetic work due to the Poynting vector ($\mathbf{E}\times \mathbf{B}/\mu_0$). 
Other (higher order) processes like 
thermal conductivity which happens at microscopic scales are not considered here. 
Finally, equation (\ref{ohm}) is the standard Ohm's law which allows us to close the system. 

The basic idea of this coronal model is the separation of the non-resolved inertial small-scale 
dynamics (see Figure \ref{Intro_Heat} and \ref{ShemaHeating}) from the large-scale dynamics  
detected with current instruments. 
{In practice, we assume that the previous MHD equations describe the large-scale behavior of the 
plasma; the nonlinear small-scale processes ({\it ie.} the effects of the inertial small-scales on the 
large-scale dynamics) will be modeled through effective dissipations, also called turbulent viscosity and 
resistivity. In other words, the usual molecular viscosity and magnetic resistivity are replaced 
by turbulent coefficients which are several orders of magnitude larger. The expression of these 
turbulent viscosities (hereafter denoted $\nu_t$ and $\eta_t$) will be derived in Section \ref{sectionSS}
and the MHD equations in the next Section.}

{An important assumption to derive this heating model is that the slow boundary motions are able 
to sustain a fully turbulent state in the corona, with a typical photospheric time longer than the 
coronal time scales. Basically, this statement means that the characteristic time of the photosphere has 
to be longer than the time needed for a perturbation to cross the coronal loops. For a typical longitudinal 
direction $\ell$ of $10^7$m and an Alfv\'en speed of several $10^6$m/s, one finds a crossing time of the 
order of 1s. For a photospheric granule with a size $\lambda=10^6$m and a velocity $\bar{u}=10^3$m/s, 
one finds a lifetime of about $10^3$s. In this case, the time scale separation is thus well satisfied. 
However, a validity condition may arise if the turbulent nature of the photospheric motion is taken 
into account since their velocities seem to follow a Kolmogorov spectrum 
(\cite{Roudier87,Chou,Espagnet}). 
Then the photospheric velocity at a given scale $\lambda$ should 
scale as $\lambda^{1/3}$ and the typical photospheric nonlinear time as $\lambda^{2/3}$, which 
decreases with the size of the granules. The associated condition is discussed and used in Section 
\ref{K41p}.}

\subsection{Large-scale solutions}

The nonlinear solutions of the previous MHD equations are non trivial and require direct numerical 
simulations on super-computers. However, we are here mainly interested in the large-scale behavior of 
the coronal flow which allows us to make some simplifications. We will seek solutions of the form 
\begin{eqnarray}
\mathbf{u} &=& u(x,z) \mathbf{e_y} , 
\label{solution1}\\
\mathbf{B} &=& B_0 \mathbf{e_z} + b(x,z) \mathbf{e_y} ,   
\label{solution2}
\end{eqnarray}
independent of $y$ and $t$, which satisfy, in particular, the divergence free conditions (\ref{divu}) and 
(\ref{divb}). These solutions do not depend on time since we assume magnetic loops maintained in 
a stationary regime of turbulence. We also observe that the fluctuating velocity and magnetic fields are 
taken perpendicular to the uniform magnetic field $\mathbf{B_0}$. This assumption is compatible with the 
fact that, in strongly magnetized media, pseudo-Alfv\'en waves are mainly along the uniform magnetic field 
whereas the nonlinear dynamics is dominated by shear-Alfv\'en waves in the transverse planes. 
We recall that shear-Alfv\'en and pseudo-Alfv\'en waves are the two kinds of linear perturbations 
about the equilibrium, the latter being the incompressible limit of slow magnetosonic waves.
Replacing the previous solution types into the MHD equations (\ref{mhd1}) and (\ref{mhd2}) leads to 
\begin{eqnarray}
\partial_x (P +b^2(x,z)/2 \mu_0) &=& 0 ,\\
\partial_z (P +b^2(x,z)/2 \mu_0) &=& 0 ,\\
B_0 \partial_z b(x,z) + \mu_0 \ \rho_0 \nu_t \Delta u(x,z) &=& 0 ,
\label{E3}\\
B_0 \partial_z u(x,z) + \eta_t \Delta b(x,z) &=& 0 .
\label{E4}
\end{eqnarray}
These equations describe the large-scale evolution of the plasma once the turbulent regime is 
established. In particular, they allow us to calculate the pressure field when the magnetic field is known. 
The linear form of equations (\ref{E3}) and (\ref{E4}) may be misleading since the small-scale 
nonlinearities are concentrated in the turbulent coefficient transports, namely the turbulent eddy 
viscosity $\nu_t$ and magnetic diffusivity $\eta_t$.  This assumption can be justified when looking at the 
nonlinear evolution equations for the energy spectra, as used in Section \ref{sectionSS}. Indeed, in Fourier 
space, a nonlocal analysis leads to diffusive effects due to the small-scale dynamics over the large scale 
motions, actually stemming from absorption terms involved  in the modeled energy transfers between 
small and large scales.

\subsection{Photospheric conditions}

The photospheric forcing introduced in equation (\ref{BoundCI_vel}) mimics the large-scale motion of 
magnetic footpoints imposed by the convective cells. Following \cite{HP92}, we assume that the frozen-in 
law is satisfied in the dense photosphere which leads, when coupled with the Ohm's law (\ref{ohm}), to 
the boundary relation for the electric field: 
\begin{eqnarray}
{\bf E} (x,z=\pm \ell)&=& \mp U(x) B_0 \mathbf{e_x} . 
\label{Efield}
\end{eqnarray}
In the close neighborhood of the photosphere, this relation is modified to take into account the magnetic 
diffusivity. This gives for the x-component: 
\begin{eqnarray}
E_x (x,z\simeq \pm \ell) &=& \mp {\eta_t} \frac{\partial b}{\partial z}(x,z) \mp U(x) B_0 .
\label{Efield2}
\end{eqnarray} 
The continuity of the tangential component of the electric field implies that 
\begin{equation}
\frac{\partial b}{\partial z}(x,z=\pm \ell) = 0.
\label{ContinuityCI_Ohm}
\end{equation} 
These photospheric conditions are of course compatible with equations (\ref{E3}) and (\ref{E4}) when 
$z=\pm \ell$.

\subsection{Energy flux injection}

The aim of this section is to express the energy flux released by photospheric motion along coronal 
loops. We are thus interested in the z-component of the flux (\ref{Feq}) at the boundaries $z=\pm \ell$, 
namely: 
\begin{equation}
\mathcal{F}_z(x,z=\pm \ell)   = \mathbf{e_z} \cdot 
\left[\mathbf{u}\cdot\left(\rho_0 \frac{u^2}{2} + P - \bar{\bar{\sigma}}\right)
+ \frac{\mathbf{E}\times\mathbf{B}}{\mu_0}\right]. 
\label{fluxF}
\end{equation}
The velocity boundary conditions (\ref{BoundCI_vel}) do not involve (convective) mass 
flows through the photospheric planes since $u_z(z=\pm l)=0$. Therefore, the two first terms in expression 
(\ref{fluxF}) do not contribute, and the input flux reduces to 
\begin{eqnarray}
\mathcal{F}_z(x,z=\pm l)&=& \Pi_z + \mathcal{P}_z,
\label{TotFlu}
\end{eqnarray}
with
\begin{eqnarray}
\Pi_z = -\left(\mathbf{u}\cdot\bar{\bar{\sigma}}\right)_z  \, \, , \, 
\mathcal{P}_z = \left(\frac{\mathbf{E}\times\mathbf{B}}{\mu_0}\right)_z,
\label{PoyntingFlu}
\end{eqnarray}
{\it i.e.} only the viscous stress tensor and the Poynting vector contribute to the energy 
flux transfer to the corona. One of the main goal of this paper is to evaluate this energy flux by first 
solving equations (\ref{E3}) and (\ref{E4}), which depend on the turbulent viscosity and resistivity, 
and second, by calculating these turbulent coefficients from Alfv\'en wave kinetic equations.

\subsection{Solving the large-scale dynamics}

In this section, we solve equations (\ref{E3}) and (\ref{E4}) to find the large-scale plasma behavior. 
We develop the x-dependence of the velocity and magnetic fields in Fourier series (since 
$h$ is assumed to be the coherence length of the magnetic and velocity fields in the x-direction) 
\begin{equation}
\xi(x,z)=\sum_{n=1}^{+\infty}\xi_n^{(1)}(z)\cos\frac{n\pi x}{h} + 
\sum_{n=1}^{+\infty}\xi_n^{(2)}(z)\sin\frac{n\pi x}{h},
\label{xifield}
\end{equation}
where $\xi$ is either $u$ or $b$. 
Note that other solutions are possible with, {\it eg.} a twisted flux tube (\cite{Inver}), which do not drastically change the coronal heating predictions. 
Substitution of (\ref{xifield}) into (\ref{E3}) and (\ref{E4}) gives
\begin{eqnarray}
B_0 \frac{d b_n^{(i)}}{dz} + \mu_0 \ \rho_0 {\nu_t} \left(\frac{d^2}{dz^2} - 
\frac{n^2\pi^2}{h^2} \right) u_n^{(i)} &=& 0,\label{eqdiffV}\\
B_0 \frac{d u_n^{(i)}}{dz} + {\eta_t} \left(\frac{d^2}{dz^2}
- \frac{n^2\pi^2}{h^2}  \right)b_n^{(i)} &=& 0 , 
\label{eqdiffB}
\end{eqnarray}
where $(i)$ stands for indices $(1)$ or $(2)$, and finally
\begin{equation}
\frac{d^4 u_n^{(i)}}{dz^4} - \left[\frac{B_0^2}{\mu_0 \rho_0 {\nu_t \eta_t}} + 
2\left(\frac{n\pi}{h}\right)^2\right]
\frac{d^2 u_n^{(i)}}{dz^2} + \left(\frac{n\pi}{h}\right)^4 u_n^{(i)} = 0. 
\label{eqdiff}
\end{equation}
Note that the magnetic field satisfies the same equation, only the boundary conditions make 
the distinction between the two fields. Using the following notation:
\begin{eqnarray}
\alpha=\frac{B_0}{\sqrt{\mu_0 \rho_0 {\nu_t \eta_t}}}, \,\;
 \alpha \lambda_n= \frac{n\pi}{h}, \, \, r^m=\frac{d^m}{d(\alpha z)^m}, 
\end{eqnarray}
equation (\ref{eqdiff}) rewrites
\begin{eqnarray}
r^4 u_n^{(i)} - (1+2\lambda_n^2)r^2 u_n^{(i)} + \lambda_n^4 u_n^{(i)} &=& 0 , 
\end{eqnarray}
whose solutions may be expressed in terms of exponentials. The photospheric boundary conditions 
(\ref{BoundCI_vel}) and (\ref{ContinuityCI_Ohm}) combined with equation (\ref{eqdiffV}) gives the 
relations
\begin{eqnarray}
u_n^{(i)}(z=\pm \ell) = \pm U_n^{(i)} \, \, \, \rm{and}  \, \,  
\frac{d^2 u_n^{(i)}}{dz^2} (z=\pm \ell) = \pm \alpha^2 \lambda_n^2 U_n^{(i)} , 
\end{eqnarray}
which lead eventually to the solutions 
\begin{equation}
u_n^{(i)}(z)=\frac{U_n^{(i)}} {\Re^+_n + \Re^-_n} \left[ \frac{\Re^-_n \sinh(\Re^+_n \alpha z)}
{\sinh(\Re^+_n \alpha \ell)} + \frac{\Re^+_n \sinh(\Re^-_n \alpha z)} {\sinh(\Re^-_n \alpha \ell)} \right],
\label{velfield}
\end{equation}
where
\begin{eqnarray}
\Re_n^{\pm}= \frac{ \sqrt{1+4\lambda_n^2} \pm 1}{2} .
\end{eqnarray}
The substitution of (\ref{velfield}) into (\ref{eqdiffV}) and (\ref{eqdiffB}) gives, after some manipulations, 
the following expression for the magnetic field Fourier coefficients 
\begin{eqnarray}
b_n^{(i)}(z) =  \frac{B_0 U_n^{(i)}} {\Re^+_n + \Re^-_n}
\left( \frac{\Re^+_n \cosh(\Re^-_n \alpha z)} {\eta\alpha\sinh(\Re^-_n \alpha \ell)} 
- \frac{\Re^-_n \cosh(\Re^+_n \alpha z)} {\eta\alpha\sinh(\Re^+_n \alpha \ell)} \right) . 
\label{magfield}                
\end{eqnarray}

\subsection{Evaluation of the energy flux}

From the solutions (\ref{velfield}) and (\ref{magfield}) it is now possible to calculate the energy 
flux (\ref{TotFlu}) at footpoint levels $z=\pm \ell$. The contribution of the Poynting 
vector to the flux is
\begin{eqnarray}
\mathcal{P}_z(x,\pm \ell)=\mp \frac{B_0}{\mu_0} U(x) b(x,z = \pm \ell) ,
\label{zz1}
\end{eqnarray}
and the contribution of the viscous stress tensor is 
\begin{eqnarray}
\Pi_z(x,\pm \ell) &=& \mp \rho_0 {\nu_t} U(x) \sigma_{zy} = 
\mp \rho_0 {\nu_t} U(x) \frac{\partial u(x,z=\pm \ell)} {\partial z} . 
\label{zz2}
\end{eqnarray}
The input flux coming from the photospheric boundaries ($z=\pm \ell$) is calculated by averaging over 
the x-periodicity of the velocity and magnetic fields (\ref{xifield}):
\begin{eqnarray}
|\mathcal{F}_z(z=\pm \ell)| = \frac{1}{2h}  \left|\int_{-h}^{+h} 
\left[\mathcal{P}_z(x,\pm \ell) + \Pi_z(x,\pm \ell) \right] dx \right| .
\end{eqnarray}
After some manipulations, we obtain the total flux (from the two footpoint levels): 
\begin{eqnarray}
|\mathcal{F}_z| &=& \rho_0 
\sum_{n=1}^\infty \sqrt{\frac{{\nu_t}}{{\eta_t}}}\frac{B_0}{\sqrt{\rho_0 \mu_0}} U_n^2
\left( \frac{\sinh\left(\alpha \ell \right)} {\cosh\left[\sqrt{1+4\lambda_n^2} 
\alpha \ell \right] - \cosh\left(\alpha \ell \right)}  \right. \nonumber\\
&+& \, \left. \frac{(1+2\lambda_n^2) \sinh\left[\sqrt{1+4\lambda_n^2} \alpha \ell \right]
/\sqrt{1+4\lambda_n^2 }} {\cosh\left[\sqrt{1+4\lambda_n^2} \alpha \ell\right] 
- \cosh\left(\alpha l\right)}  \right) \, ,
\label{FluSec11}
\end{eqnarray}
with $U_n^2={U^{(1)}}^2_n+{U^{(2)}}^2_n$. 
{Note the appearance of a turbulent magnetic Prandtl number ($\nu_t/\eta_t$) whose origin 
is directly linked to the action of nonlinear small-scale processes on large-scales. We will later 
show that this number may be taken equal to unity in the Alfv\'en wave turbulence regime. 
Also note that the turbulent viscosities enter in expression (\ref{FluSec11}) through the 
coefficients $\alpha$ and $\lambda_n$. The origin of the presence of $\nu_t$ and $\eta_t$ is thus 
due, on the one hand, to the large-scale MHD equations (\ref{E3})--(\ref{E4}) and, on the other hand, 
to the Poynting vector (\ref{zz1}) and the viscous stress tensor (\ref{zz2}) which also include the 
small-scale nonlinear retroaction.}
The energy injected from the magnetic footpoints is propagated through the corona by Alfv\'en waves. 
Then, the collisions between "upward" and "downward" wave packets lead to nonlinear dynamics 
which is characterized by a turbulent cascade that transfers energy from the energy injection scales 
down to the smallest dissipative scales where energy is destroyed, and the corona heated.
The coronal loops reach a stationary state of fully developed turbulence when the dissipation 
equals the electromagnetic and viscous stress.

\subsection{Turbulent spectrum of photospheric velocities}
\label{K41p}

The coronal heating flux is now evaluated from (\ref{FluSec11}). 
The  turbulent viscosities entering into this flux expression will be computed in the next Section.
Therefore, the last unknown is the magnitude of the velocity at the photospheric boundaries 
(namely $U_n$). 
The goal of this section is to evaluate the scale dependence of photospheric velocities.

As reported by different measurements (\cite{Roudier87,Chou,Espagnet}), the  photospheric 
velocities are assumed to follow a Kolmogorov isotropic spectrum
\begin{eqnarray} 
u^2(k)= \frac{2}{3}\bar{u}^2 k_{inj}^{2/3} k^{-5/3},
\label{specPhoto}
\end{eqnarray}
with 
\begin{eqnarray} 
\bar{u}^2=\int_{k_{inj}}^{+\infty} u^2(k) dk = \sum_{n=1}^{+\infty} U_n^2/2 \ ,
\label{relation1}
\end{eqnarray}
where $\bar{u}$ is the photospheric rms velocity, $U_n^2$ is the power spectrum of the boundary 
velocity fluctuations, and 
$k_{inj}$ is the largest scale at which energy is injected into the system; it corresponds to the size of 
the largest granular cells, {\it i.e.} about $1000$km. In other words, it is the integral scale of the 
turbulent photospheric flow from which the inertial range starts. The wave number $k_{inj}$ may be 
directly connected to the thickness of the arcade, such that $k_{inj} =\pi/h$. More generally, the 
wave number $k$ and the harmonic number $n$ may be connected through the relation $k=n \pi/h$. 
Converting the sum in equation (\ref{relation1}) into an integral gives the relation
\begin{eqnarray} 
U_n^2 = \frac{2\pi}{h} u^2(k) = \frac{4}{3}\bar{u}^2n^{-5/3}.
\label{photospectrum}
\end{eqnarray}

The basic idea of the model is that Alfv\'en wave packets interact nonlinearly along coronal magnetic 
structures in which the plasma is assumed to be maintained in a fully turbulent state because of 
the slow photospheric motion. We have seen in Section \ref{mhdmodel} that a condition may arise 
on time-scales since the photosphere is also turbulent and its typical nonlinear time, $\tau^{ph}_{nl}$, 
may be very short for small granules. In order to satisfy this condition, we impose that 
$\tau^{ph}_{nl}(\lambda)> \tau_{cross}$, where $\tau_{cross}$ is the time needed for a perturbation 
to cross half of the loops of length $2\ell$ at speed $B_0/\sqrt{\rho_0 \mu_0}$, namely
\begin{eqnarray} 
\tau_{cross} \sim \frac{\ell \sqrt{\rho_0 \mu_0}}{B_0} \ ,
\end{eqnarray}
and $\tau^{ph}_{nl}(\lambda)$ is the eddy turnover (nonlinear) time of granular structures 
at a typical scale $\lambda$ and velocity $u(\lambda)$;
\begin{eqnarray} 
\tau^{ph}_{nl}(\lambda) \sim \frac{\lambda}{u(\lambda)}
\sim\frac{\lambda^{2/3}}{\bar{u}} \left( \frac{2\pi}{k_{inj}}\right)^{1/3} .
\end{eqnarray}
This time-scale separation induces for the wave number $k=2\pi/\lambda$, the inequality
$k<k_{cross}$, where 
\begin{eqnarray} 
k_{cross} =\frac{\pi}{h}N_{cross}, \,\; \rm{with} \,\;
N_{cross}=\left(2\frac{h}{\ell} \frac{B_0/\sqrt{\rho_0 \mu_0}}{\bar{u}}\right)^{3/2} .
\label{AlfvNumber}
\end{eqnarray}
Hence, the upper limit of the summation on harmonic numbers in (\ref{FluSec11}) is given by 
$N_{cross}$ which is a large number since the Alfv\'en velocity is much larger than the 
photospheric rms velocity. For that reason, the expression (\ref{FluSec11}) may also be written 
as an integral with 
\begin{eqnarray} 
U^2(k) = \frac{4}{3} \left(\frac{\pi}{h}\right)^{5/3} \bar{u}^2 k^{-5/3} .
\end{eqnarray}
The energy injection produced by the photospheric motion is finally rewritten as follows: 
\begin{eqnarray}
|\mathcal{F}_z|=
\rho_0 \int_{k_{inj}}^{k_{cross}} \sqrt{\frac{\nu_t}{{\eta_t}}} \frac{B_0}{\sqrt{\rho_0 \mu_0}} 
U^2(k) \Psi(k) \frac{h}{\pi} \, dk,
\label{FluSec12}
\end{eqnarray}
with 
\begin{eqnarray}
\Psi(k)&=& \frac{\sinh(\alpha \ell)} {\cosh\left[\sqrt{\alpha^2+4k^2}\ell\right]-
\cosh(\alpha \ell)}\nonumber \\
&+& \frac{\alpha^2+2k^2}{\alpha \sqrt{\alpha^2+4k^2}} 
\frac{\sinh\left[\sqrt{\alpha^2+4k^2}\ell\right]}
{\cosh\left[\sqrt{\alpha^2+4k^2}\ell\right]-\cosh(\alpha \ell)}.
\label{PhiFunction}
\end{eqnarray}
Under the time-scale condition discussed above, the coronal structure is submitted to slow 
external forcing. Note that other time scales are present in our problem; namely, the nonlinear coronal 
time, $\tau_{nl}$, and the Alfv\'en time, $\tau_A$, which has nothing to do with the crossing time 
$\tau_{cross}$ introduced above. It may be interpreted as the time of interaction between two 
counter-propagating Alfv\'en wave packets. Both times will be used in Section \ref{sectionSS} where 
the nonlinear small-scale dynamics will be analyzed. 
The main assumption made in such wave turbulence dynamics is that the Alfv\'en time is smaller than 
the nonlinear coronal time.
Note that in MHD the transfer time-scale is built on $\tau_{nl}$ and $\tau_A$ which makes a great 
difference in Navier-Stokes fluids for which the transfer time is the nonlinear time. This point 
will be further discussed in the next section.  
Then the following hierarchy of times may be estimated 
\begin{eqnarray}
\tau_A < \tau_{nl} < \tau_{cross} < \tau^{ph}_{nl} \, .
\end{eqnarray}
As seen in Section \ref{mhdmodel}, the nonlinear photospheric time can be as large as 
$10^3$s and the crossing time is about $1$s. 
For the two other characteristic time scales, no observational constraints can be used. 
However, recent direct numerical simulations (\cite{Bigot08b}) show clearly that a turbulent 
MHD flow evolving in a strong uniform magnetic field ($B_0>b$) exhibits a time ratio 
$\tau_A  / \tau_{nl}<1$, which decreases as the strength of the uniform field increases.

\section{Small-scale dynamics and anisotropy}
\label{sectionSS}
{In the previous Section, we have been able to evaluate the flux of energy released by photospheric 
motion along coronal loops. The calculation has been greatly simplified by the assumption of a scale 
separation between large and small (inertial) scales, and thus the introduction of turbulent {eddy 
diffusivities ($\nu_t$ and $\eta_t$). In fact, all the complexity of the nonlinear dynamics is concentrated in these coefficients.
We will now try to evaluate them by going back to previous works on Alfv\'en wave turbulence. 
We first summarise previous results on anisotropic MHD turbulence. 
We then introduce the asymptotic equations of Alfv\'en wave turbulence which are not rederived here (see \cite{Galtier00}), and show their main properties. 
Finally, we derive the turbulent viscosities from such nonlinear equations.}

\subsection{Role of anisotropy in MHD turbulence}

The first description of incompressible MHD turbulence proposed by Iroshnikov and Kraichnan (IK) 
in the 1960s (\cite{Iro64, Kraich65}) is based on a modified version of the  Kolmogorov dimensional 
analysis where the large-scale magnetic effects are taken into account although the fluid is supposed 
to be globally isotropic. The IK's heuristic model includes Alfv\'en waves which approximate the 
magnetic field at the largest scales as a uniform magnetic field. Then, it is the sporadic and  
successive
collisions between counterpropagating Alfv\'en waves that produce the turbulent cascade. At leading 
order, the nonlinear dynamics is driven by three wave interactions, and the transfer time through the 
scales is estimated by $\tau_{tr} \sim \tau_{nl}^2/\tau_{A}$, where $\tau_{nl}\sim \lambda/u_{\lambda}$ 
is the nonlinear turnover time whereas $\tau_{A}\sim \lambda/ {\tilde B_0}$ is the Alfv\'en wave period 
(where $\tilde B_0$ stands for the Alfv\'en speed, taken as the rms magnetic field fluctuation). 
Hence, the energy spectrum is $k^{-3/2}$. 

The weakness of the IK phenomenology is the apparent contradiction between the presence of Alfv\'en 
waves and the absence of a strong uniform magnetic field. This point and the role of a strong external 
magnetic field $\mathbf{B_0}$ have been widely discussed in the literature, in particular  
during the last two decades (Strauss 1976, Montgomery \& Turner 1981; Shebalin et al. 1983; 
Oughton et al. 1994; Goldreich \& Sridhar 1995; Ng \& Bhattacharjee 1996; Matthaeus et al. 1998; 
Galtier et al. 2000; Nazarenko et al., 2001; Milano et al. 2001; 
Chandran 2005; Bigot et al. 2008a). One of the most clearly established results is the bi-dimensionalization 
of a turbulent flow with a strong reduction of the nonlinear transfers along the $\mathbf{B_0}$ ambiant field. 

In the case of coronal structures, like magnetic arcades or open magnetic fields in coronal holes, a strong 
magnetic field at large-scale is clearly present. This field is often modeled as a uniform field through, for 
example, the approximation of reduced MHD (see, {\it e.g.}, Dmitruk \& Matthaeus 2003; Buchlin et al. 
2007). It was thought until recently (see {\it eg.} \cite{Oughton04}) that the reduced MHD approximation 
was only able to describe the regime of strong turbulence but a recent work (\cite{naza}) has extended 
its domain of application to the wave turbulence regime as well. This point is particularly important for 
our model which is based on the latter regime. In our model, the magnetic arcade is composed of a 
network of parallel strands that are maintained in a fully developed turbulent regime. 
Contrary to the isotropy hypothesis originally assumed by \cite{HP92}, we include anisotropic effects 
through turbulent viscosities. In this approach, it is thus necessary to distinguish between the 
perpendicular ($\perp$) and parallel ($\parallel$) directions to the uniform $\mathbf{B_0}$ field.

An important anisotropic property discussed in the literature (Higdon 1984; Goldreich \& Sridhar 1995; 
Galtier et al. 2005; Boldyrev 2006) is the interdependence of perpendicular and parallel scales. 
According to direct numerical simulations (see, {\it e.g.}, Cho \& Vishniac 2000; Maron \& Goldreich 2001; 
Shaikh \& Zank 2007; Bigot et al. 2008b), one of the most fundamental results seems to be the (critical) 
balance between the nonlinear and Alfv\'en times which leads to the scaling relation 
$k_\parallel \sim k^{2/3}_\perp$. In other words, a turbulent MHD flow evolving in a strongly magnetized 
medium seems to be characterized by an approximately constant ratio (generally smaller than one) 
between the Alfv\'en and the nonlinear times. This result (generally) implies a dynamics mainly driven 
by Alfv\'en wave interactions.

\subsection{Alfv\'en wave turbulence}

The incompressible MHD equations (\ref{mhd1})--(\ref{divb}) may be reformulated in terms of
the Els\"asser fluctuations 
\begin{eqnarray}
\mathcal{\mathbf{z}}^s=\mathbf{u} + s \mathbf{\tilde b} \, , 
\end{eqnarray}
with magnetic fields written in velocity units ($\mathbf{\tilde b} = \mathbf{b} / \sqrt{\mu_0 \rho_0}$), 
{and in the inviscid case ({{\it ie.} in the absence of} the dissipative terms proportional to the kinematic 
viscosity and the magnetic diffusivity)}, 
\begin{eqnarray}
\partial_t \mathcal{\mathbf{z}}^s - s \mathbf{\tilde B_0} \cdot \mathbf{\nabla} \mathcal{\mathbf{z}}^s
&=& - \mathcal{\mathbf{z}}^{-s}\cdot\mathbf{\nabla}  \mathcal{\mathbf{z}}^s - \mathbf{\nabla} P_* \, , \\
\nabla \cdot {\bf z}^s &=& 0 \, .
\label{ZMHDeq}
\end{eqnarray}
Here, $s=\pm$, stands for the directional polarity, indicating the propagating direction of waves along 
$\mathbf{\tilde B_0}=\mathbf{B_0} / \sqrt{\mu_0 \rho_0}$; $P_*$ is the total (kinetic plus magnetic) 
pressure. 
{Note that { the present analysis is focused on the nonlinear plasma dynamics from which the 
turbulent transport coefficients will be derived.}}
In the presence of a strong uniform magnetic field, 
MHD turbulence may be dominated by wave dynamics for which the nonlinearities are weak, and 
thus $\tau_{A} \ll \tau_{nl}$. In this limit, a small formal parameter $\epsilon$ can be introduced to measure 
the strength of the nonlinearities to give, for the jth-component 
\begin{eqnarray}
\left(\partial_t -s {\tilde B_0} \partial_\parallel \right)z_j^s 
= -\epsilon z^{-s}_m {\partial_m} z^s_j - {\partial_j} P_* \  
\label{etf}
\end{eqnarray}
(with Einstein's notation used for indices).
Note that the parallel direction ($\parallel$) corresponds to the z-direction. 

{We will Fourier transform such equations, {with the following definitions for 
the Fourier transform of the Els\"asser field components $z^s_j(\mathbf{x},t)$:} 
\begin{eqnarray}
z^s_j({\bf x},t) = \int a^s_j(\mathbf{k},t) \, e^{i(\mathbf{k} \cdot {\bf x}+s \omega_k t)} \, d\mathbf{k} \, ,
\end{eqnarray}
where $\omega_k={\tilde B_0} k_\parallel$ is the Alfv\'en frequency. The quantity $a^s_j(\mathbf{k},t)$ 
is the wave amplitude in the interaction representation, hence the factor $e^{is\omega_k t}$. The 
Fourier transform of equation (\ref{etf}) thus is  
\begin{eqnarray}
\partial_t a^s_j(\mathbf{k})=-i\epsilon k_m P_{jn} \int a^{-s}_m(\pmbmath{\kappa}) a^s_n(\mathbf{L}) 
e^{is\Delta\omega t} \delta_{\mathbf{k},\pmbmath{\kappa}\mathbf{L}}
d\pmbmath{\kappa} \, d\mathbf{L} \ .
\label{amplitude}
\end{eqnarray}
Here, $P_{jn}$ is the projector on solenoidal vectors such that $P_{jn}(\textbf{k})=\delta_{jn}-k_j k_n/k^2$; 
$\delta_{\mathbf{k},\mathbf{\kappa}\mathbf{L}}=\delta(\mathbf{k}-\pmbmath{\kappa}-\mathbf{L})$ reflects 
the triadic interaction, and $\Delta\omega=\omega_L-\omega_k-\omega_\kappa$ is the frequency mixing.
The appearance of an integration over wave vectors $\pmbmath{\kappa}$ and $\mathbf{L}$ is directly 
linked to the quadratic nonlinearity of equation 
(\ref{etf}) {(as a result of the Fourier transform of a correlation product)}.}

Equation (\ref{amplitude}) {is the {compact expression of the} incompressible MHD equations when a strong uniform magnetic is present.} 
It is the point of departure of the wave turbulence formalism which consists of writing equations for the 
long time behavior of second order moments. In such an 
{asymptotic and statistical} development, the time-scale 
separation, $\tau_{A}/\tau_{nl} \ll 1$, leads to the destruction of some nonlinear terms, including the fourth 
order cumulants, and only the resonance terms survive (\cite{Galtier00,Galtier02}). It allows one to obtain a 
natural asymptotic closure for the moment equations. 
{In such a statistical development, the following general definition for the total (shear- plus 
pseudo-Alfv\'en wave) energy spectrum is used;
\begin{equation}
\langle a^s_j(\mathbf{k}) a^s_j(\mathbf{k^{\prime}}) \rangle = 
E^s(\mathbf{k}) \, \delta(\mathbf{k}+\mathbf{k^{\prime}}) / k_\perp \, ,
\label{esh0}
\end{equation}
where $\langle \rangle$ stands for ensemble average and $k_\perp=\vert \mathbf{k_\perp} \vert$.}
In absence of helicities {and in the case of an axially symmetric turbulence,
the asymptotic  equations simplify. For the shear-Alfv\'en waves, the energy spectrum is given by  
\begin{equation}
E_{shear}^s(k_\perp,k_\parallel)=f(k_\parallel) E^s_\perp(k_\perp) \, ,
\label{esh}
\end{equation}
where $f(k_\parallel)$ is a function fixed by the initial conditions ({\it ie.} there is no energy transfer along 
the parallel direction). The transverse part obeys the following nonlinear equation (the small parameter 
$\epsilon$ is now included in the time variable)}
\begin{eqnarray}
\partial_t E^s_\perp(k_\perp)=
\frac{\pi}{\tilde B_0} \int \int \cos^2\phi \sin\theta  \, \frac{k_\perp}{\kappa_\perp}
E^{-s}_\perp(\kappa_\perp) \nonumber\\
\left[k_\perp E^s_\perp(L_\perp)-L_\perp E^s_\perp(k_\perp)\right]
d\kappa_\perp dL_\perp , 
\label{shearEq}
\end{eqnarray}
where $\phi$ is the angle between $\mathbf{k_\perp}$ and $\mathbf{L_\perp}$, and $\theta$ is the angle 
between $\mathbf{k_\perp}$ and $\pmbmath{\kappa}_\perp$ {with the perpendicular wave vectors 
satisfying the triangular relation $\mathbf{k_\perp} = \mathbf{L_\perp} + \pmbmath{\kappa}_\perp$ 
(see Figure \ref{triad}). 
{Note that from the axisymmetric assumption}, the azimuthal angle integration has already
been performed, and we are only left with an integration over the absolute values of the
two wave numbers, $\kappa_\perp = |\pmbmath{\kappa}_\perp|$ and $L_\perp = |\mathbf{L_\perp}|$.}
In the same way, equations can be written for pseudo-Alfv\'en waves which are 
passively advected by shear-Alfv\'en waves, namely 
\begin{eqnarray}
\partial_t E^s_\parallel(k_\perp) = \frac{\pi}{\tilde B_0}\int \int
\sin\theta \, \frac{k_\perp}{\kappa_\perp} E^{-s}_\perp(\kappa_\perp) \nonumber\\
\left[k_\perp E^s_\parallel(L_\perp) -L_\perp E^s_\parallel(k_\perp)\right] d\kappa_\perp dL_\perp ,
\label{pseuEq}
\end{eqnarray}
{with by definition
\begin{equation}
E_{pseudo}^s(k_\perp,k_\parallel)={\tilde f}(k_\parallel) E^s_\parallel(k_\perp) \, ,
\end{equation}
where ${\tilde f}(k_\parallel)$ is a function fixed by the initial condition.}
{Equations (\ref{shearEq}) and (\ref{pseuEq}) -- called wave kinetic equations -- 
only involve perpendicular nonlinear dynamics (with 
a dependence only on the perpendicular components of the wave vectors), 
a situation expected from previous works (Strauss 1976, Montgomery \& Turner 1981; Shebalin et al. 1983; 
Oughton et al. 1994; Goldreich \& Sridhar 1995; Ng \& Bhattacharjee 1996; Matthaeus et al. 1998) 
{that showed the decrease of the parallel nonlinear transfers with the increasing strength of the external magnetic field 
$\mathbf{\tilde B_0}$}.}
\begin{figure}[ht]
\centering
\includegraphics[width=8.cm]{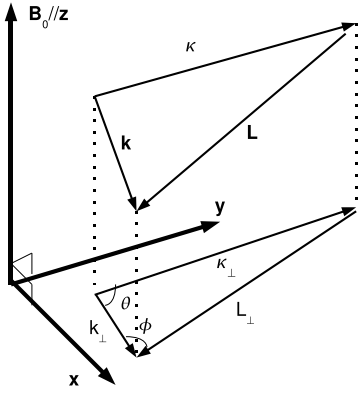}
\caption{Triadic interaction $\mathbf{k}=\pmbmath{\kappa} + \mathbf{L}$ and its projection in the plane 
perpendicular to $\mathbf{B}_0$ (such that $\mathbf{k}_\perp=\pmbmath{\kappa}_\perp + \mathbf{L}_\perp$).}
\label{triad}
\end{figure}

\subsection{Anisotropic turbulent viscosities}

The main goal of this Section is the derivation of turbulent viscosities in the context of strongly 
anisotropic MHD turbulence. We have seen that wave turbulence leads to the inhibition of nonlinear 
transfers in the $\mathbf{\tilde B_0}$ direction, and to the bidimensional reduction of the dynamics. 
This result makes a strong difference to purely isotropic MHD turbulence from which 
the turbulent viscosities were derived in the original model (\cite{HP92}). 
We believe that taking into account the flow anisotropy is an important improvement 
in the description of the coronal heating problem. 

The wave kinetic equations are assumed to describe the nonlinear dynamics at small inertial 
scales. These scales correspond to the scales unresolved by the current spacecrafts which are 
reported in Figure \ref{Intro_Heat}. The turbulent viscosities represent the average effect 
of nonlinearities at these small scales, and they appear in the large-scale equations 
(\ref{E3})--(\ref{E4}). In practice, we make a Taylor expansion of the wave kinetic equations for 
nonlocal triadic interactions (\cite{Pouq76}) that satisfy the relation  
\begin{eqnarray}
k_\perp \ll \kappa_\perp, L_\perp, 
\end{eqnarray}
which leads to 
\begin{eqnarray}
L_\perp \simeq \kappa_\perp -k_\perp \cos \theta, 
\label{51}
\end{eqnarray}
for triadic interactions $\mathbf{k}_\perp=\pmbmath{\kappa}_\perp+\mathbf{L}_\perp$ (see Figure \ref{triad}). 
Then, relation (\ref{51}) is substituted into the kinetic equations (\ref{shearEq})-(\ref{pseuEq}) for, 
respectively, the shear-Alfv\'en waves 
\begin{eqnarray}
\partial_t E^s_\perp(k_\perp)=- k^2_\perp E^s_\perp(k_\perp)\nonumber \\ 
\int \frac{\pi}{\tilde B_0} E^{-s}_\perp(\kappa_\perp) d\kappa_\perp 
\int_0^{2\pi} \cos^2\theta \sin^2\theta \, d\theta,
\end{eqnarray}
and the pseudo-Alfv\'en waves
\begin{eqnarray}
\partial_t E^s_\parallel(k_\perp)=
- k^2_\perp E^s_\parallel(k_\perp)\nonumber \\
\int \frac{\pi}{\tilde B_0} E^{-s}_\perp(\kappa_\perp) d\kappa_\perp  
\int_0^{2\pi} \sin^2\theta \, d\theta.
\end{eqnarray}
Finally, we obtain:
\begin{eqnarray}
\partial_t E^s_{\perp}(k_\perp) &=& -\nu^s_\perp k_\perp^2 E^s_{\perp}(k_\perp) ,\\
\partial_t E^s_{\parallel}(k_\perp) &=& -\nu^s_\parallel k_\perp^2 E^s_\parallel(k_\perp),
\end{eqnarray}
where $\nu_\perp^s$ and $\nu_\parallel^s$ are the {so-called} turbulent viscosities, corresponding respectively 
to shear- and pseudo-Alfv\'en waves, defined as
\begin{eqnarray}
\nu_\perp^s &=& \frac{\pi^2}{4 \tilde B_0} \int E^{-s}_{\perp}(\kappa_\perp) \, d\kappa_\perp,
\label{ShearVisco}\\
\nu_\parallel^s &=& \frac{\pi^2}{\tilde B_0} \int E^{-s}_{\perp}(\kappa_\perp) \, d\kappa_\perp .
\label{PseuVisco} 
\end{eqnarray}
{The turbulent viscosities are thus directly obtained after integration over the $\theta$ angle.
Note that, similarly to the original paper (Heyvaerts \& Priest 1992), the integration is made from the 
beginning of the inertial range (at $\kappa_{inj}$) where the injection of energy is made; this choice 
contrasts with some models where a wavenumber cut-off is chosen inside the inertial range which gives 
a k-dependence of the turbulence viscosity (see {\it eg.} B\ae renzung et al. 2008).}
An evaluation of the turbulent viscosities is now possible by the use of the exact power law solutions 
of the integro-differential equations (\ref{shearEq}) and (\ref{pseuEq}). This is first made in the 
context of a magnetic arcade for which the inward and outward Alfv\'en waves are exactly balanced  
(balanced case). This approach is then generalized to open magnetic fields 
for which outward Alfv\'en waves are dominant (unbalanced case).

\subsection{Balanced turbulence for a magnetic arcade}
\label{BTMA}
For balanced turbulence, as said before, it is not necessary to distinguish outward 
from inward Alfv\'en waves. We thus drop the directional polarity $s$, and we obtain
\begin{eqnarray}
\nu_\perp &=& \frac{\pi^2}{4 \tilde B_0} \int E_{\perp}(\kappa_\perp) \, d\kappa_\perp,
\label{ShearViscoB}\\
\nu_\parallel &=& \frac{\pi^2}{\tilde B_0} \int E_{\perp}(\kappa_\perp) \, d\kappa_\perp .
\label{PseuViscoB} 
\end{eqnarray}
In this case, the exact power law solution of equation (\ref{shearEq}) is (\cite{Galtier00})
\begin{eqnarray}
E_\perp(k_\perp)=C_K {\tilde B_0}^{1/2}P_\perp^{1/2}k_\perp^{-2}, 
\label{spek2}
\end{eqnarray}
where $C_K$ is the Kolmogorov constant whose value is about $0.585$, and $P_\perp$ is the 
constant energy flux of shear-Alfv\'en waves {whose definition is
\begin{eqnarray}
\partial_t E_\perp(k_\perp) = -\partial_{k_\perp} P_\perp(k_\perp) \, .
\label{fluxx}
\end{eqnarray}
It is important to note that it is the condition of constant energy flux that allows to find (by the 
Zakharov transformation) the exact power law solution of the nonlinear wave turbulence equation.}
The introduction of (\ref{spek2}) into (\ref{ShearViscoB}) and (\ref{PseuViscoB}) gives 
\begin{eqnarray}
\nu_{\perp} &=& \frac{\pi^2}{4}\sqrt{\frac{P_\perp}{\tilde B_0}} \frac{C_K}{\kappa_{inj}}, \\
\nu_{\parallel} &=& \pi^2 \sqrt{\frac{P_\perp}{\tilde B_0}} \frac{C_K}{\kappa_{inj}}. 
\label{TurbVisco}
\end{eqnarray}
Note the relation $\nu_{\parallel} = 4\nu_{\perp}$ between the turbulent 
coefficients corresponding to pseudo- and shear-Alfv\'en waves. In particular, this 
means that the pseudo-Alfv\'en wave energy is susceptible to decay faster than the shear-Alfv\'en 
one. Actually, this property has been detected recently in numerical simulations of freely decaying
MHD turbulence (\cite{Bigot08a}). 
Finally note that in this wave turbulence regime, there is an automatic equipartition state between 
the kinetic and magnetic energies. Therefore, no distinction will be made between the average 
nonlinear effects acting on the velocity and on the magnetic field. A unit turbulent magnetic 
Prandtl number will then be taken ({\it ie.} $\nu_t/\eta_t=1$).

\section{Magnetic arcade heating}
\label{chauffage1}
\subsection{Connection between large- and small-scale dynamics}

The magnetic arcade being supposed  in a fully turbulent state, the energy transfer rate is then 
assumed to be constant at each scales in the inertial range, from the largest photospheric scales 
at which energy is injected, down to {the small dissipative scale fom which heating starts to occur. 
This fundamental remark allows us to link the first estimate of the energy flux 
(\ref{FluSec12}), based on the photospheric motions, to the flux through the anisotropic arcade 
scales, estimated from the turbulent eddy viscosity and magnetic diffusivity. }

{According to relations (\ref{esh}) and (\ref{fluxx}) the energy flux $P_\perp$, introduced above, 
satisfies the general relation 
\begin{eqnarray}
\partial_t E_{shear}(k_\perp,k_\parallel) = -f(k_\parallel) \, \partial_{k_\perp} P_\perp . 
\end{eqnarray}
We remind that} $f$ is an arbitrary dimensionless function which reflects the absence of 
nonlinear transfers along the $\mathbf{\tilde B_0}$ direction. The coronal heating in a fraction of the 
arcade, say $Q$, can thus be estimated from the small-scale dynamics by
\begin{eqnarray}
Q = 2 \ell \, 2h \, L \, \rho_0 P_\perp \int f(k_\parallel) dk_\parallel , 
\end{eqnarray}
where $L$ delimits a fraction of the arcade thickness in the y-direction and $Q$ is measured in 
Js$^{-1}$ (Note the presence of the constant mass density). In the same manner, we may evaluate the 
coronal heating from the large-scale energy flux ${|\mathcal{F}_z|}$. It gives for the same coronal 
volume:
\begin{eqnarray}
Q = 2h \,L\, |\mathcal{F}_z| . 
\end{eqnarray}
Hence, the relation between both photospheric and turbulent fluxes:
\begin{eqnarray}
P_\perp = \frac{|\mathcal{F}_z|} {2\ell \rho_0 \int f(k_\parallel)dk_\parallel}.
\label{PfctF}
\end{eqnarray}

\subsection{General expression of the heating flux}

The small-scale analysis gives two turbulent viscosities coming from, respectively, the shear-
and pseudo-Alfv\'en wave energy. To be consistent with our model, we will only consider the 
former one since, in the strongly anisotropic limit of wave turbulence, the pseudo-Alfv\'en waves 
fluctuate only in the parallel direction to the ambiant magnetic field. 
However, we note that the inclusion of the pseudo waves effect modifies only slightly 
the form of the turbulent viscosity (by a factor $5$). Then, relation (\ref{PfctF}) gives 
(with $\nu_t \equiv \nu_{\perp}$)
\begin{eqnarray}
\frac{16}{\pi^4} {\nu_t}^2 {\tilde B_0} \frac{\kappa_{inj}^2}{C_K^2} = 
\frac{|\mathcal{F}_z|} {2\ell \rho_0 \int f(k_\parallel)dk_\parallel}. 
\end{eqnarray}
Together with the injection wave number $\kappa_{inj} = \pi / h$, related to the arcade thickness,
we obtain 
\begin{eqnarray}
|\mathcal{F}_z| = \frac{32}{\pi^2 C_K^2} \frac{{\nu_t}^2}{h^2} \ell
\rho_0 {\tilde B_0} \int f(k_\parallel)dk_\parallel. 
\label{TurbFlu}
\end{eqnarray}
The substitution of (\ref{FluSec12}) into (\ref{TurbFlu}) leads to the relation
\begin{eqnarray}
\frac{32}{\pi} \frac{{\nu_t}^2 \ell}{C_K^2 h^3} \int f(k_\parallel)dk_\parallel = 
\int_{k_{inj}}^{k_{cross}} U^2(k) \Psi(k) \, dk ,
\label{equalFlu}
\end{eqnarray}
where a unit turbulent magnetic Prandtl number is taken
($\nu_t=\eta_t$). The above expression also gives 
\begin{eqnarray}
\Lambda_0 \,\alpha^2 \int^{k_{cross}}_{k_{inj}} \Psi(k) k^{-5/3} dk=1 ,
\label{EqInt}
\end{eqnarray}
where
\begin{eqnarray}
\Lambda_0=\frac{\pi C_K^2}{24} \left(\frac{\pi}{h}\right)^{5/3} \frac{h^3 \bar{u}^2}{\ell {\tilde B_0}^2}
\frac{1}{\int f(k_\parallel) dk_\parallel} . 
\end{eqnarray}
In the next section, we use the following approximation:
\begin{eqnarray}
\int f(k_\parallel) \, dk_\parallel = \pi/\ell.
\end{eqnarray}

\subsection{Different scale contributions}

According to the {specific} value of the $k$ wave number, different estimates may be made for the function 
$\Psi(k)$ that enters in the expression (\ref{FluSec12}) of the energy injection by photospheric 
velocities, namely
\begin{eqnarray}
\alpha \ll k &:&\Psi_1(k) \simeq \frac{k}{\alpha} , \\
\sqrt{\frac{\alpha}{\ell}} \ll k \ll \alpha &:& \Psi_2(k) \simeq 2 , \\
k \ll \sqrt{\frac{\alpha}{\ell}} &:&\Psi_3(k) \simeq \frac{\alpha}{k^2 \ell} .
\end{eqnarray}
{We recall that $\alpha=B_0 / \sqrt{\mu_0 \rho_0 \nu_t \eta_t}={\tilde B_0} / \nu_t$; it is 
therefore a way to measure, for example, the relative importance of the diffusive terms in equations 
(\ref{E3}) and (\ref{E4}) which can be rewritten as
\begin{eqnarray}
\alpha \partial_z {\tilde b(x,z)} + \Delta u(x,z) &=& 0 ,
\label{E3b}\\
\alpha \partial_z u(x,z) + \Delta {\tilde b(x,z)} &=& 0 .
\label{E4b}
\end{eqnarray}
On can also more easily evaluate the relative importance of the Poynting vector (\ref{zz1}) and the 
stress tensor (\ref{zz2}) to the flux with the relation
\begin{eqnarray}
\mathcal{P}_z(x,\pm \ell) + \Pi_z(x,\pm \ell)  = \\
\mp \rho_0 U(x) \nu_t [ \alpha {\tilde b}(x,z = \pm \ell) + \partial_z u(x,z=\pm \ell) ] \nonumber .
\end{eqnarray}
For the case $\alpha \ll k$, it is clear that the main contribution to the flux comes from the stress 
tensor, whereas for the two other $k$ inequalities, the main contribution comes from the Poynting 
vector.}
Relation (\ref{EqInt}) can thus be written in these different regimes as: 
\begin{eqnarray}
(i) \, \, \alpha \ll k_{inj}, \nonumber
\end{eqnarray}
\begin{eqnarray}
\Lambda_0 \alpha^2 \int_{k_{inj}}^{k_{cross}} \Psi_1(k) k^{-5/3} dk = 1;
\label{ipp1}
\end{eqnarray}
\begin{eqnarray}
(ii) \, \, k_{inj} \ll \alpha \ll k_{inj}^2 \ell , \nonumber
\end{eqnarray}
\begin{eqnarray}
\Lambda_0 \alpha^2 \left(\int_{k_{inj}}^{\alpha} \Psi_2(k) k^{-5/3} dk + 
\int_{\alpha}^{k_{cross}} \Psi_1(k) k^{-5/3} dk \right) = 1 ;
\label{ipp2}
\end{eqnarray}
\begin{eqnarray}
(iii) \,  \, k_{inj}^2 \ell \ll \alpha \ll k_{cross} , \nonumber
\end{eqnarray}
\begin{eqnarray}
\Lambda_0 \alpha^2 \left(\int_{k_{inj}}^{\sqrt{\alpha/\ell}} \Psi_3(k) k^{-5/3} dk
+ \int_{\sqrt{\alpha/\ell}}^{\alpha} \Psi_2(k) k^{-5/3} dk\right. \nonumber\\
+ \left.\int_{\alpha}^{k_{cross}} \Psi_1(k) k^{-5/3} dk\right) = 1 ;
\label{ipp3}
\end{eqnarray}
\begin{eqnarray}
(iv) \,  \, k_{cross} \ll \alpha \ll k_{cross}^2\ell , \nonumber
\end{eqnarray}
\begin{eqnarray}
\Lambda_0 \alpha^2 \left( \int_{k_{inj}}^{\sqrt{\alpha/\ell}} \Psi_3(k) k^{-5/3} dk 
+ \int_{\sqrt{\alpha/\ell}}^{k_{cross}} \Psi_2(k) k^{-5/3} dk \right) = 1 ;
\label{ipp4}
\end{eqnarray}
\begin{eqnarray}
(v) \,  \, k_{cross}^2 \ell \ll \alpha , \nonumber
\end{eqnarray}
\begin{eqnarray}
\Lambda_0 \alpha^2 \int_{k_{inj}}^{k_{cross}} \Psi_3(\alpha,k) k^{-5/3} dk = 1 .
\label{ipp5}
\end{eqnarray}
These relations simplify respectively to:
\begin{eqnarray}
(i) \,  \, 3 \Lambda_0 \alpha \left(k_{cross}^{1/3} - k_{inj}^{1/3} \right) =1 ;
\end{eqnarray}
\begin{eqnarray}
(ii) \,  \, 3\Lambda_0 \left(\alpha k_{cross}^{1/3} + \alpha^2 k_{inj}^{-2/3} - 2\alpha^{4/3} \right)=1 ;
\end{eqnarray}
\begin{eqnarray}
(iii) \,  \, 3\Lambda_0 \left(\frac{\alpha^3}{8\ell}k_{inj}^{-8/3} + \frac{7}{8}\alpha^{5/3} \ell^{1/3}
+ \alpha k_{cross}^{1/3} -2\alpha^{4/3} \right)=1 ;
\end{eqnarray}
\begin{eqnarray}
(iv) \,  \, 3\Lambda_0 \left(\frac{\alpha^3}{8\ell}k_{inj}^{-8/3} - \alpha^2 k_{cross}^{-2/3} 
+ \frac{7}{8} \alpha^{5/3} \ell^{1/3} \right)=1 ;
\label{forcorona}
\end{eqnarray}
\begin{eqnarray}
(v) \,  \, \frac{3}{8}\Lambda_0 \frac{\alpha^3}{\ell} \left( k_{inj}^{-8/3} - k_{cross}^{-8/3} \right)=1.
\end{eqnarray}
These five different regimes correspond to a different intensity of the small-scale nonlinearities. 
This intensity is measured in terms of an equivalent wave number $\alpha$ and it is compared to the 
two available scales, namely $k_{inj}$ and $k_{cross}$. Case (i) is the situation where the small-scale  
nonlinearities are the strongest: they are felt beyond $k_{inj}$. Case (v) is the situation where the 
small-scale nonlinearities are the weakest: the diffusive terms in equations (\ref{E3b})--(\ref{E4b}) are 
then negligible.

\subsection{Predictions for a coronal plasma}

We define the following basic coronal quantities: 
\begin{eqnarray}
h &=& 10^6 \, \mathcal{H} \, \, \rm{m} ,\\
\ell &=& 10^7 \, \mathcal{L} \, \, \rm{m} ,\\
B_0 &=& 10^{-2} \, \mathcal{B}_0 \, \, \rm{T} ,\\
{\bar u} &=& 10^3 \, \mathcal{U} \, \, \rm{m \, s^{-1}} ,\\
\rho_0 &=& 10^{-12} \, \mathcal{M} \, \, \rm{kg \, m^{-3}} ,
\end{eqnarray}
which give the estimates: 
\begin{eqnarray}
{{\tilde B_0}} &=& 8.9 \times 10^6 \, \mathcal{B}_0 \, \mathcal{M}^{-1/2} \, \, \rm{m \, s^{-1}} ,\\
k_{inj} &=& 3.1 \times 10^{-6} \, \mathcal{H}^{-1} \, \, \rm{m^{-1}} ,\\
k_{cross} &=& 2.4 \times 10^{-4} \, \mathcal{H}^{1/2} \, \mathcal{L}^{3/2} \,
\mathcal{U}^{3/2} \, \mathcal{B}_0^{-3/2} \, \mathcal{M}^{-3/4} \, \, \rm{m^{-1}} ,\\
\Lambda_0 &=& 1.2 \times 10^{-1} \, \mathcal{H}^{4/3} \, \mathcal{U}^2 \,
\mathcal{B}_0^{-2} \, \mathcal{M} \, \, \rm{m}^{4/3} .
\end{eqnarray}
With these values, the only satisfied equation is (\ref{forcorona}) which corresponds to 
the inequality of case (iv). The solution of equation (\ref{forcorona}) is:
\begin{equation}
\alpha = 4.1 \times 10^{-3} \, \mathcal{H}^{-4/3} \, \mathcal{B}_0^{2/3} \, \mathcal{M}^{-1/3} \, \,  
\mathcal{U}^{-2/3} \, \, \rm{m^{-1}} \, .
\end{equation}
Then, it gives the following estimate for the turbulent viscosity:
\begin{eqnarray}
{\nu_t} = 2.2 \times 10^9 \, \mathcal{H}^{4/3} \, \mathcal{U}^{2/3} \, \mathcal{B}_0^{1/3} \, 
\mathcal{M}^{-1/6} \, \, \rm{m^2 s^{-1}} .
\end{eqnarray}
Hence, the heating flux per unit area
\begin{eqnarray}
|\mathcal{F}_z| = 1249  \, \mathcal{H}^{2/3} \, \mathcal{U}^{4/3} \, \mathcal{B}_0^{5/3} \, 
\mathcal{M}^{1/6} \, \, \rm{J \, m^{-2} s^{-1}} .
\label{balanceflux}
\end{eqnarray}
The turbulent velocity of the coronal plasma may be found from relation
\begin{eqnarray}
u_\perp^2 = \frac{\pi}{h}  \int^{+\infty}_{k_{inj}} C_K {\tilde B_0}^{1/2} P_\perp^{1/2}k_\perp^{-2} \, dk_\perp .
\end{eqnarray}
Only the perpendicular fluctuating velocity is taken into account since we are only concerned 
with shear-Alfv\'en waves in a strongly anisotropic turbulence. We substitute $P_\perp$ by 
its expression (\ref{PfctF}) to finally obtain the {following} prediction:
\begin{eqnarray}
u_{\perp} = 50  \, \mathcal{H}^{1/2} \, \mathcal{U}^{1/3} \, \mathcal{B}_0^{2/3} \, 
\mathcal{M}^{-1/6} \,  \mathcal{L}^{-1/2} \, \, \rm{km \, s^{-1}} .
\end{eqnarray}

These results compare favorably with observations, in particular, in the quiet corona with a heating 
flux large enough to explain the observations. Note that the heating prediction is mostly sensitive to 
the magnetic field intensity $\mathcal{B}_0$ which has the larger power law index by $5/3$. Thus, 
$\mathcal{B}_0=3$--$4$ ($B_0=3$--$4 \times 10^{-2}\,\rm{G}$) leads to a factor about ten {times} larger for the 
heating flux with a value close to the measurements $10^4 \rm{J \, m^{-2} s^{-1}}$ for active regions. 

\section{Coronal hole heating}
\label{chauffage2}
\subsection{Geometry of open magnetic field lines}

In this Section, the model prediction for the heating rate is extended to coronal holes where the plasma 
is guided along a large-scale magnetic field which expands into the interplanetary medium. It is 
along such structures (mainly at the poles) that the fast solar wind is released whereas the slow wind 
is freed at lower latitudes around the equatorial plane. In these configurations of open magnetic lines, 
the reflection of outward Alfv\'en waves, due to some inhomogeneities, produces inward waves 
which eventually sustain nonlinear interactions. In this case, the coronal heating is clearly 
dependent on the reflection rate of Alfv\'en waves (\cite{Velli93}, \cite{Dmitruk01}). 
The precise origin of the partial reflection of Alfv\'en waves is still under debate. 
Therefore, any theoretical prediction is useful for the comparison between observations and 
models, and eventually for the understanding of the solar corona dynamics. In this Section, we seek theoretical predictions like the reflection rate needed to sustain an efficient coronal heating localized 
in coronal holes. We see that a small reflection rate of $15-20 \%$ is enough to recover the coronal 
heating observations. This result may constrain the efficiency of the mechanisms invoked to produce 
reflected Alfv\'en waves.

\begin{figure}[ht]
\includegraphics[width=8.5cm]{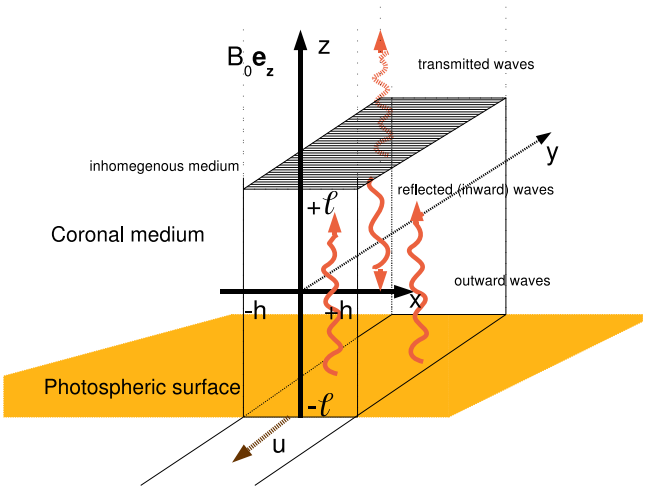}
\caption{Coronal hole geometry: the coronal plasma is confined in a volume delimited by $-h<x<+h$ 
and $-\infty<y<+\infty$, and may escape from the low coronal medium at the top of the structure 
(for $z>\ell$). The photosphere appears as a (single) lower boundary plane at altitude $z=-\ell$. 
The interplanetary inhomogeneity allows the reflection of a fraction of outward Alfv\'en waves which 
produces inward waves.}
\label{OpenLineModel}
\end{figure}
Figure \ref{OpenLineModel} shows a schematic view of open magnetic field lines which only differs from 
the magnetic loop configuration (Figure \ref{LoopModel}) is the upper boundary condition. This second 
photospheric surface is replaced by a permeable boundary from which outward Alfv\'en waves may be 
partially reflected. Note that the first model of magnetic arcades corresponds to the case where outward 
Alfv\'en waves are totally reflected and where outward and inward waves are balanced.

\subsection{Reflection rate versus cross-helicity}

To quantify the reflection rate of Alfv\'en waves (also called the Alfv\'enicity), a characteristic quantity 
is the normalized cross-helicity $\sigma_c$ defined as 
\begin{eqnarray}
\sigma_c=\frac{{z^+}^2-{z^-}^2}{{z^+}^2+{z^-}^2} \, ,
\label{corr}
\end{eqnarray}
where $z^+$ is the Alfv\'en inward wave and $z^-$ is the outward wave (with $z^s=\vert {\bf z^s} \vert$, 
$s=\pm$).
Such a quantity provides a measure of the relative importance {between}  outgoing and ingoing Alfv\'en waves. 
In particular, $\sigma_c=\pm 1$ involves the presence of only one type of polarity, and consequently, the absence 
of nonlinear interactions (thus no heating at dissipative scales), whereas a balance between $z^+$ and 
$z^-$ waves implies $\sigma_c= 0$ leading to more important nonlinear interactions (between waves of 
different polarity) as was assumed in the magnetic arcade (loop) configuration 
{analyzed in previous Sections.} 
The reduced cross-helicity is indeed related to the wave reflection rate, $\mathcal{R}$, {such as}
\begin{eqnarray}
\mathcal{R}= \frac{1- \vert \sigma_c \vert} {1+ \vert \sigma_c \vert} \, .
\end{eqnarray}
The case $\mathcal{R}=1$ (all outward waves are reflected) is similar to the loop configuration 
in which there is the same number of inward and outward waves, whereas 
in the case of $\mathcal{R}=0$ no waves are reflected.

\subsection{Unbalanced turbulence for coronal holes}

The integro-differential kinetic equations (\ref{shearEq}) for the Alfv\'en waves  are numerically 
integrated with a logarithmic subdivision of the k axis : $k_i=\delta k 2^{i/F}$, where $i$ is an integer 
($i=1, N$), $\delta k=0.125$ is the smallest wave number reached in the computation, and $F=8$ is 
the refinement of the grid. (This method was previously performed and optimized in \cite{Galtier00}.) 
This numerical technique allows to reach larger Reynolds numbers since, for a wave number 
resolution $N=157$, the maximum wave number reached, $k_{max}$, is around $10^5$ (the 
magnetic Prandtl number is equal to unity). Note that dissipative terms are introduced in the 
inviscid kinetic equations to avoid numerical instabilities, and the viscosity is fixed to 
$10^{-5}$. The initial Els\"asser fields are 
injected in the wave number range $[\delta k,44]$ with an energy spectrum proportional to 
$k_\perp^3\exp\left(-k^2_\perp/2\right)$, peaking at $k_{\perp_0}\sim 1$. 
The corresponding $U_{rms}$ and integral scale $L_0$ are, respectively, about $4.75$ and $1.5$,
which gives an initial Reynolds number ($U_{rms}L_0/\nu$) of about $10^6$. 
The flow is then left to freely evolve. 

In Figure \ref{spectrum80}, we show {instantaneous} energy spectra of the 
shear-Alfv\'en waves, $E_\perp^+$ and $E_\perp^-$, obtained for, 
respectively, $z_\perp^+$ and $z_\perp^-$ fluctuations with a cross-helicity of $0.8$ 
($\mathcal{R}=25\%$) (In the following we respectively consider $z_\perp^-$ and $z_\perp^+$ 
as the inward (reflected) and outward waves). 
\begin{figure}[ht]
\centering
\includegraphics[width=8.5cm]{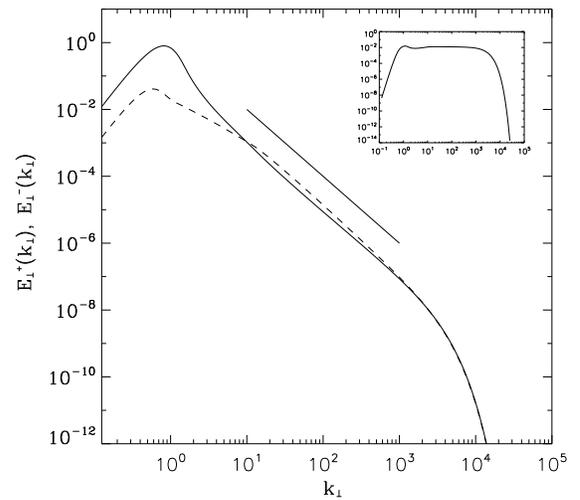}
\caption{Energy spectra of shear-Alfv\'en waves $E_\perp^+$ (solid line) and $E_\perp^-$ 
(dashed line) with $\mathcal{R}=25\%$ ($\sigma_c=0.8$). The straight line follows 
a $k_\perp^{-2}$ law which is the theoretical prediction for the balanced turbulence case 
($\mathcal{R}=100\%$ and $\sigma_c=0$).
Inset: compensated product of energy spectra 
as $E_\perp^+\,E_\perp^-\,k_\perp^4$.}
\label{spectrum80}
\end{figure}
The chosen time is the one at which the energy spectra are the most developed, {\it ie.} the most extended towards large wave numbers $k_\perp$. Different behaviors are clearly found for the 
$E_\perp^+$ and $E_\perp^-$ spectra with, at large scales, a domination of inward Alfv\'en waves 
linked to the choice of the initial conditions. 
At smaller scales, we see the appearance of an extended inertial range where the inward 
waves ($E_\perp^-$) slightly dominate. Note that the energy 
spectra product, $E_\perp^+ E_\perp^-$, follows a $k_\perp^{-4}$ scaling law as 
predicted {theoretically} (\cite{Galtier00}), and shown in the inset of Figure \ref{spectrum80}.
Finally the two energy spectra overlap at dissipative scales.
From such a numerical simulation, we are able to compute the turbulent viscosities for unbalanced 
turbulence by integration of the spectra over wave numbers, from the 
beginning of the inertial range up to the dissipative scales. This computation is done 
for several simulations corresponding to different reflection rates.

\subsection{Parametric study and predictions}

A parametric study of the heating flux according to the reflection rate of outward waves is performed. 
The turbulent viscosities are calculated for different values of the reflection rate by integrating the 
shear-Alfv\'en wave spectra, from the injection scale (normalized to unity in the numerical 
simulation) up to the dissipative scales, and by using {relation (\ref{ShearVisco}) to estimate $\nu_\perp^s$.}
{In such a calculation, the chosen time is the one at which the energy spectra are most developed.} 
Since the result depends on the initial amount of energy taken in the simulation, the flux is 
normalized to the flux obtained in the loop case (see Section \ref{chauffage1}) for which a balanced 
turbulence is assumed. With this method, we are able to approximately predict 
the variation of the relative heating flux as a function of the reflection rate. 

From the turbulent flux expression (\ref{TurbFlu}), we obtain the relations 
\begin{eqnarray}
\frac{|\mathcal{F}^s_z(\mathcal{R})|}{|\mathcal{F}_z|} \simeq
\left( \frac{{\nu^s}_\perp(\mathcal{R})}{\nu_\perp} \right)^2 = 
\frac{\left[\int E^{-s}_{\perp}(\mathcal{R}) \, d\kappa_\perp\right]^2}
     {\left[\int E_{\perp} \, d\kappa_\perp\right]^2} \, ,
\end{eqnarray}
which depend on the reflection rate of outward waves $\mathcal{R}$. 
$\mathcal{F}_z^s$ and $\mathcal{F}_z$ are, respectively, the heating fluxes in the unbalanced and 
balanced turbulent cases. The presence of the directional 
polarity $s=\pm$ keeps track of the advection between waves of different polarity. 
The calculation made here concerns the heating flux at the lower 
boundary ($z=-\ell$) since the upper boundary condition previously used is no longer the same.
We have also assumed, in particular, that the Kolmogorov constant 
does not change drastically for different reflection rates. 
Figure \ref{Alfvenicity} shows the flux ratio as a function of the reflection rate $\mathcal{R}$. 
As expected, the curve decreases with the reflection rate, meaning that the turbulent heating is 
{clearly less efficient for smaller values of $\mathcal{R}$.} 
Finally, the flux ratio goes to zero when 
only one type of wave remains (since in this case the nonlinear interactions disappear). 
Note that the decreasing function is close to a linear curve with a flux ratio of about $40\%$ for 
$\mathcal{R}=50\%$. 
\begin{figure}[h]
\centering
\includegraphics[width=8.5cm]{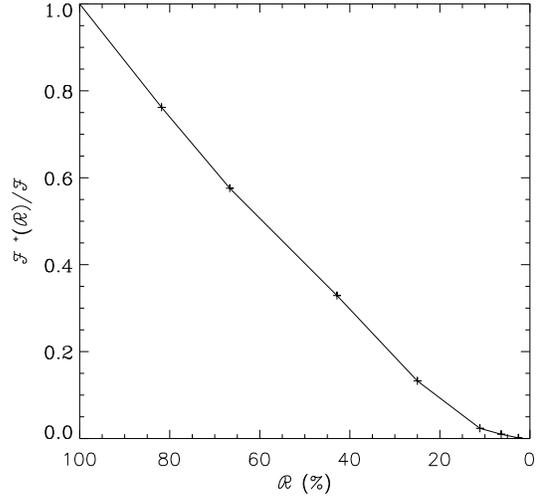}
\caption{Evolution of the heating flux according to the reflection rate of outward Alfv\'en waves. 
The heating flux is normalized to the flux obtained in the balanced turbulence case (Section 
\ref{chauffage1}), {\it i.e.} for $\mathcal{R}=100\%$.}
\label{Alfvenicity}
\end{figure}

The heating flux prediction in coronal holes is known and estimated as around 
$100\,\rm{J \, m^{-2} s^{-1}}$ (\cite{Withbroe77}). Since the background magnetic field is still 
about $10^{-2}$T, we may find the reflection rate in coronal holes by using both the relation 
(\ref{balanceflux}) and the curve in Figure \ref{Alfvenicity}. The value $\mathcal{R}=25\%$ 
leads to a flux ratio of about $15\%$--$10\%$, a small enough value to recover the heating 
flux for coronal holes (without taking into account the fast solar wind). We see that a relatively 
small reflection rate is therefore enough to produce an efficient coronal heating due to anisotropic 
MHD turbulence.

\section{Discussion and conclusion}
\label{end}
In this paper, we have developed an analytic model for strongly anisotropic structures  
in order to recover well known coronal heating rates for the quiet sun, active regions 
and coronal holes. The coronal structures are assumed to be in a turbulent state maintained 
by the slow motion of the magnetic footpoints anchored on the photospheric surface. 
The main difficulty is that existing spacecraft are unable to resolve all the inertial (small) scales, 
and {\it a fortiori} the dissipative scales. 
So, firstly, the injected energy at large scale is resolved from incompressible 
MHD equations, and for specific boundary conditions satisfying the divergence free condition. 
Secondly, the (unresolved) small-scale dynamics is modeled by turbulent viscosities 
derived from an asymptotic (exact) closure model of wave turbulence for the case of the loop 
configuration (also called balanced turbulence), 
{\it i.e.} when there are as many inward as outward waves which nonlinearly interact. 
In the open magnetic line configuration the nonlinear interactions are sustained by the conversion of 
outward waves into inward waves by reflection (unbalanced turbulence). We have numerically 
integrated the kinetic equation of Alfv\'en waves for different values of the reflection rate, and  
plotted the evolution of the coronal heating flux according to this rate, as a measure of 
the relative magnitude of ingoing and outgoing waves. 
Finally these results are compared to the balanced turbulence flux (loop {configuration}). 

For standard loop geometry parameters, and for a magnetic field intensity of $B_0 = 10^{-2} \rm{T}$, 
we find a heating flux prediction of about  $1.2 \times 10^3 \rm{J \, m^{-2} s^{-1}}$ 
which is close to the value measured in the quiet sun ($10^3 \rm{J \, m^{-2} s^{-1}}$). 
Moreover, the prediction for the turbulent velocity ($50 \, \rm{km \, s^{-1}}$)   
compares favorably  with the measurements of nonthermal velocities ($30 \, \rm{km \, s^{-1}}$) 
in the quiet solar corona, as well as with the maximal values of some line profiles 
($55 \, \rm{km \, s^{-1}}$) found by the SUMER instrument (\cite{Chae98}).
For active regions, our estimation tends toward a value of about $10^4\, \rm{J \, m^{-2} s^{-1}}$ 
for a magnetic field intensity stronger ($3$--$4$ times larger) than for the quiet sun, which is 
an expected value for strong solar activity. 
For the heating rate prediction in open magnetic lines, we take as a reference the loop configuration 
(for which there is a balanced turbulence) and the same parameters as for the quiet sun. 
We can recover the well known estimates for coronal holes of about $100\, \rm{J \, m^{-2} s^{-1}}$
(without taking into account the fast solar wind) for a reflection rate of outward waves of about $15$--$20\%$.

The previous predictions do not take into account the coronal heating by pseudo-Alfv\'en waves
{(nor the heating due to the {pure} 2D state { - with wave vectors such as 
(${\bf k}_\perp, k_\parallel=0)$ -}
 which is not described by Alfv\'en wave turbulence)}, 
since their parallel fluctuations are excluded by the boundary conditions (involving only 
perpendicular fluctuations). 
Thus, their insertion should heat slightly more the solar corona and raise our predictions slightly. 
Moreover, in the computation of heating flux for the unbalanced turbulence case, only the heating due to 
outward Alfv\'en waves is considered. The heating due to inward waves should modify slightly 
the heating prediction for coronal holes, and therefore allows a slight decrease of the reflection rate with 
a heating rate maintained around at $100\, \rm{J \, m^{-2} s^{-1}}$.

\begin{acknowledgements}
Financial support from PNST/INSU/CNRS are gratefully acknowledged. This work was supported by 
the ANR project no. 06-BLAN-0363-01 ``HiSpeedPIV''. We thank the anonymous referee for 
comments and questions which have improved the presentation of the paper.
\end{acknowledgements}

\end{document}